\pdfoutput=1
\documentclass{JINST}

\usepackage{amsmath}
\usepackage{textcomp} 
\usepackage{ifpdf}

\DeclareGraphicsExtensions{.pdf,.png,.jpg} 

\title{NIKEL\_AMC: Readout electronics for the NIKA2 experiment}
\author{O.~Bourrion$^a$\thanks{Corresponding author.}, 
A.~Benoit$^b$,
J.~L.~Bouly$^a$,
J.~Bouvier$^a$,
G.~Bosson$^a$,
M.~Calvo$^b$,
A.~Catalano$^a$,
J.~Goupy$^b$,
C.~Li$^a$,
J.~F.~Mac\'ias-P\'erez$^a$,
A.~Monfardini$^b$,
D.~Tourres$^a$,
N.~Ponchant$^a$,
C.~Vescovi$^a$.\\
\llap{$^a$}LPSC, Universit\'e Grenoble-Alpes, CNRS/IN2P3 \\
53, rue des Martyrs, Grenoble, France \\
\llap{$^b$}Institut N\'eel, CNRS, Universit\'e Grenoble-Alpes \\
  25, rue des Martyrs, Grenoble, France \\
}  

\abstract{
The New Iram Kid Arrays-2 (NIKA2) instrument has recently been installed at the IRAM 30 m telescope. NIKA2 is a state-of-art instrument dedicated to mm-wave astronomy using microwave kinetic inductance detectors (KID) as sensors. The three arrays installed in the camera, two at 1.25 mm and one at 2.05 mm,  feature a total of 3300 KIDs.
To instrument these large array of detectors, a specifically designed electronics, composed of 20 readout boards and hosted in three microTCA crates, has been developed.
The implemented solution and the achieved performances are presented in this paper. We find that multiplexing factors of up to 400 detectors per board can be achieved with
homogeneous performance across boards in real observing conditions, and a factor of  more than 3 decrease in volume with respect to previous generations.
} 

\keywords{Instruments for CMB observations; Electronic detector readout concepts; Data acquisition concepts}

\begin{document}
\section{Introduction}
\label{intro}
Kinetic inductance detectors (KID) are innovative superconducting detectors, typically operated below 200\,mK.
They consist of high-quality superconducting resonant circuits electromagnetically coupled to a transmission line.
They are designed to resonate in the microwave domain \cite{Baselmans,Day,DoyleThesis}. 
These detectors were used in the New Iram Kid Arrays (NIKA) instrument, \cite{Monfardini,Monfardini2011,Monfardini2014} which was installed at the 30-meter telescope of the Institut de Radioastronomie Millim\'etrique (IRAM).
In that instrument, two KID arrays allowed simultaneous sky observations at two frequency bands: 150 and 260\,GHz.
The arrays, featuring respectively 132 and 224 Hilbert type Lumped Element KID (LEKID), were simultaneously instrumented by the New Iram Kid ELectronics (NIKEL) \cite{Bourrion2012}.
Several technical observation campaigns showed that NIKA was producing state of the art scientific results \cite{Catalano,Adam0,Adam1,Adam2,Ritacco}.
As a consequence, it was used by external astronomers during three  more observational campaigns.

NIKA2, the next generation KID camera,  has been recently installed in the IRAM 30 m telescope, hence becoming the first ever KID-based resident camera in the world.
It should stay in place for a period of at least 10 years.
It will be offering the possibility to study a large variety of astrophysical and cosmological subjects going from the Sunyaev-Zeldovich effect in
clusters of galaxies to star formation in our galaxy. NIKA2 is equipped with three KID arrays: one array with 1020 pixels for imaging the sky at 150\,GHz and two arrays of 1140 pixels each for imaging the sky 260\,GHz with polarimetry. This constitutes not only a major step forward from the instrumental point of view leading to a factor of  5-9 more detectors per array, but also from the astrophysical point of view with a factor of 10 improvement in the mapping speed thanks to the full coverage of the telescope field-of-view. 

Although the NIKEL readout electronic performances were fully satisfactory for NIKA, the factor of 10 increase in the number of detectors for NIKA2 forced us to upgrade the electronics preserving their state-of-art performance. This upgrade is related to three main issues. First, the power consumption had to be decreased by a factor of two in order to avoid system issues at the telescope like for example excessive power dissipation in the telescope receiver cabin and overconsumption with respect to the maximum available power. Second, the total volume occupied by the readout system had to be decreased by a factor of three to fit in the telescope's receiver cabin. Third, the system maintainability had to be enhanced to guarantee a large uptime, corresponding at least to the NIKA2 resident time. These requirements lead us to develop crate adapted compact electronic boards integrating both the radio frequency and digital subsystems. 

NIKA and NIKA2 are instruments unique in their kind, at the forefront of the recent KID technology; all the detection chain, from the front-end to the back-end described here are pioneer of this promising technology.
This paper, which describes the readout electronics used for NIKA2, is organized as follows: section~\ref{InstruMetho} presents the system,
section~\ref{elecDescr} describes the readout electronics and eventually section~\ref{sysPerfSec} presents the performances achieved.
\section{System description}
\label{InstruMetho}
The setup used to instrument a KID array feed-line and its associated electronics is extensively described in \cite{Bourrion2011}. 
In summary, the excitation frequency comb is generated at baseband in the electronics using coordinate rotation digital computer (CORDIC), up-converted with an in-phase/quadrature (IQ) mixer to the 1.3 to 2.4\,GHz KID frequencies and injected in the resonator line. 
The returning, and thus modified, frequency comb is down-converted and analyzed by channelized Digital Down Converters (DDC) that provides In-phase (I) and Quadrature (Q) components.
These components are used to determine each tone amplitude and phase.

The KID arrays used for NIKA2 were produced \cite{Goupy} to meet the bandwidth specification of the NIKEL electronics \cite{Bourrion2012}, i.e. 500\,MHz.
Fabrication dispersion can cause the resonance frequencies of two KIDs to be closer than desired, which leads to an unacceptably high level of cross-talk. To limit the impact of this effect, an average frequency separation of $\sim$4\,MHz was chosen for the 260\,GHz array (resonances from 1.9\,GHz to 2.4\,GHz) and of $\sim$2\,MHz for the 150\,GHz array (resonances from 1.3\,GHz to 1.8\,GHz).
Consequently, the 150\,GHz is instrumented by 4 feed-lines having around 250 KIDs per feed-line and the 260\,GHz arrays are each instrumented by 8 feed-lines having about 140 KIDs per feed-line.
Thus, the three NIKA2 KID arrays, composed of 3300 resonators in total, are composed of 20 resonator lines requiring 20 readout electronics.

\begin{figure}
\begin{center}
\includegraphics[angle=-0,width=0.7\textwidth]{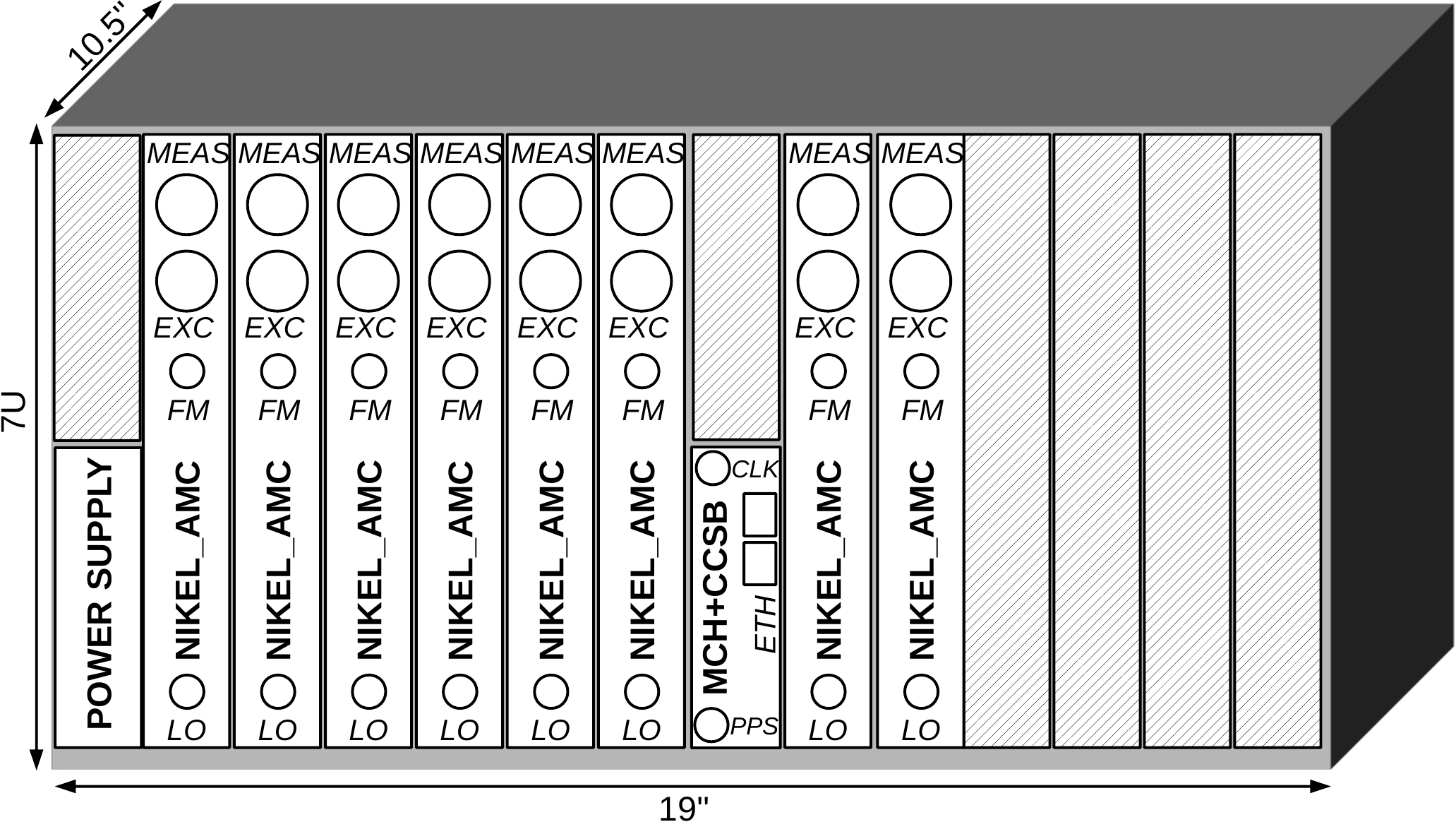}
\caption{Overview of one array readout electronics crate.
It is equipped with 4 (or 8) readout boards lodged in Advanced Mezzanine Card slots (NIKEL\_AMC), one central clocking and synchronization board (CCSB) mounted on the MicroTCA Carrier Hub (MCH) and one 600\,W power supply.
The crate allocated to the 150\,GHz channel uses 4 NIKEL\_AMC boards while the others use 8 NIKEL\_AMC boards.\label{crateFig}}
\end{center}
\end{figure}

Following the requirements detailed in section~\ref{intro}, the readout electronics were redesigned to fit in three micro Telecom Computer Architecture (microTCA) crates \cite{MTCA,VT891}: one per array.
Each crate features twelve full-size, double width Advanced Mezzanine Card (AMC) slots.
As shown in  fig.~\ref{crateFig}, each crate is equipped with 4 (or 8) readout boards lodged in Advanced Mezzanine Card slots (NIKEL\_AMC), one central clocking and synchronization board (CCSB) mounted on the MicroTCA Carrier Hub (MCH) and one 600\,W power supply.
The crate allocated to the 150\,GHz channel uses 4 NIKEL\_AMC boards while the others use 8 NIKEL\_AMC boards.

The CCSB is installed on a commercial MCH from NAT\textregistered \cite{NAT} and is used to distribute the 10\,MHz reference clock (CLK) and the Pulse Per Second (PPS) via the crate backplane.
The reference clock is generated by a FS725 Rubidium frequency standard (Stanford Research Systems\textregistered) and converted into a TTL level square signal by a custom clock distribution box and fed to the three crates.
The PPS is generated by a Global Positioning System (GPS) receiver installed at IRAM that is distributed to all instruments to permit synchronization and avoid long-term frequency drift with respect to the telescope pointing.

The MCH, which hosts the CCSB, plays a central role.
As required by the microTCA specification \cite{MTCA}, it is in charge of managing the crate (slot activation for hot-plug, power supply monitoring, fan speed controlling, sensors monitoring) and hosts a Gigabit Ethernet (GbE) hub function for communicating with each slot and the MCH extension.
This latter functionality is used for configuring and reading-out the NIKEL\_AMC and the CCSB.

Each NIKEL\_AMC includes the digital electronics as well as the Radio-Frequency (RF) chains required to instrument one feed-line of up to 400 KIDs.
The KID excitation signal is provided through the excitation output (EXC), passed through the feed-line hosted by the KID array located in the cryostat and returned to the measurement input (MEAS).
The Local Oscillator (LO) used to up and down-convert the frequency comb is generated by Aeroflex/IFR/Marconi 2024 signal generator, dedicated to the crate.
Its output is distributed to each NIKEL\_AMC, thanks to a passive commercial RF splitter. 
To minimize the RF phase noise and avoid long-term shifts, the synthesizer is also referenced by CLK.
A key point in operating with LEKID is to convert the observed in phase (I(t)) and in quadrature (Q(t)) signal to absorbed optical power.
For ground-based experiments, the resonance frequency of each KID may change substantially during an observation as a result of variations in the sky emission, which impacts photometric reproducibility.
As described in \cite{Calvo}, we deal with this issue by incorporating a system by imposing a several kHz modulation on the local oscillator frequency; this modulation is synchronous with the Field Programmable Gate Array (FPGA) readout.
\section{KID readout board}
\label{elecDescr}
\subsection{Hardware description}
\begin{figure}
\begin{center}
\includegraphics[angle=0,width=0.9\textwidth]{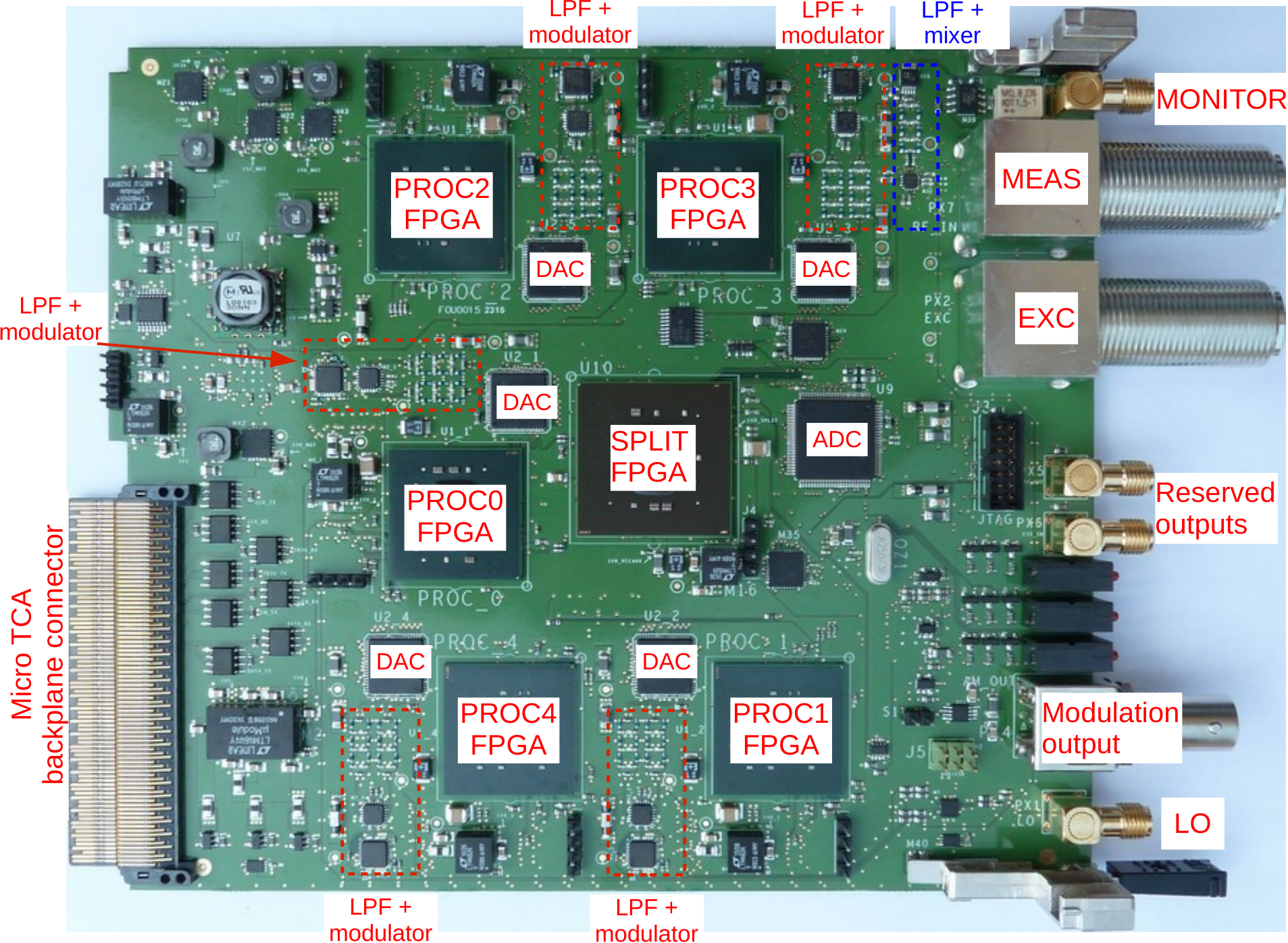}
\caption{NIKEL\_AMC board picture.
It has dimensions of $\rm 18 \times 15 \times  3\,cm^3$.
\label{PictureNikelAMC}}
\end{center}
\end{figure}

The NIKEL\_AMC boards reuse, with some improvements, the digital architecture that was developed for the previous readout electronics \cite{Bourrion2012}.
A NIKEL\_AMC board is able to manage up to 400 resonators over a bandwidth of 500\,MHz in a frequency domain ranging from 1\,GHz to 3\,GHz (limited by the RF component only, i.e. mixers, modulators, amplifiers, ...).
A picture of the board can be seen in figure~\ref{PictureNikelAMC}.
It is a 12-layer printed circuit board (PCB) having dimensions of $\rm 18 \times 15 \times  3\,cm^3$.

\begin{figure}
\begin{center}
\includegraphics[angle=0,width=1\textwidth]{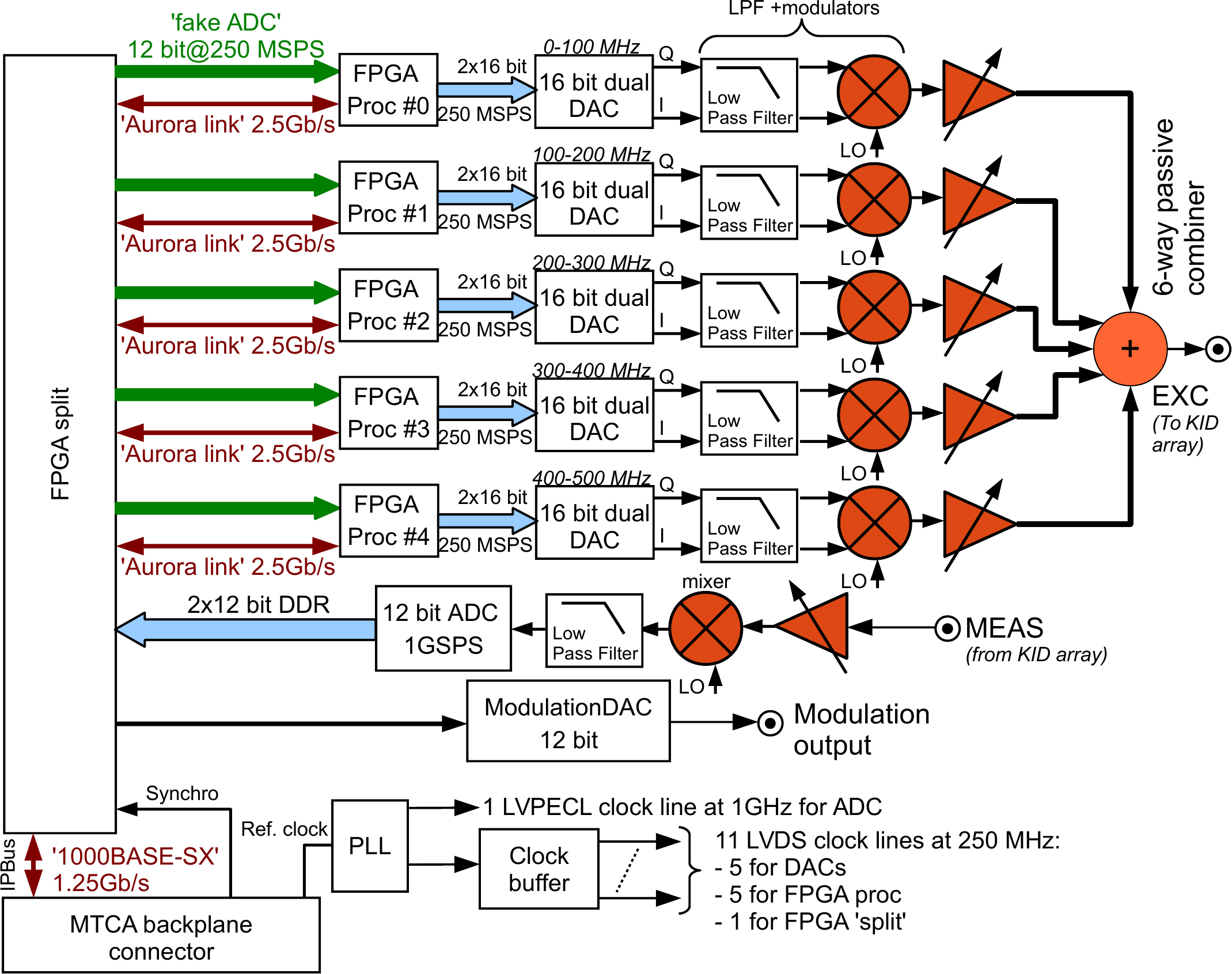}
\caption{NIKEL\_AMC board overview.
Radio-frequency parts are colored in orange.
\label{NIKEL_AMC_Hw}}
\end{center}
\end{figure}

\subsubsection{Digital part description}
To take advantage of the Digital to Analog Converter (DAC) integrated modulating and interpolating capabilities (AD9125 from analog devices), the  NIKEL\_AMC is designed to manage five adjacent bands of 100\,MHz. 
Each DAC has its carrier frequency programmed at a convenient value (respectively for each band 0\,Hz, 109.375\,MHz, 203.125\,MHz, 296.125\,MHz and 406.25\,MHz) and its internal interpolation filters selected accordingly to filter the appropriate band of signal.
As shown in fig.~\ref{NIKEL_AMC_Hw}, the digital part of the electronic board is composed of a central FPGA (labeled `split') and of five processing FPGAs (labeled `proc').
Each of the latter is driving its associated DAC with the adequate frequency comb which can feature up to 80 tones.

The `split' FPGA receives the data produced by the 12 bit ADC (ADS5400 from Texas Instruments) operated at 1 Giga Samples Per Second (GSPS).
The ADC data, that cover a bandwidth of 500\,MHz, are processed with a 5 output polyphase filter to produce five adjacent frequency bands.
Each of these frequency bands is covering a bandwidth of a 100\,MHz.

The `split' FPGA is connected to each of the five `proc' FPGA with two points to point proprietary links.
The first of these, labeled `fake ADC', is a 12 bit parallel Low Voltage Differential Signal (LVDS) link running at 250\,MSPS.
It is used to transport the polyphase filter output to the `proc FPGA' generating the excitation signal in the corresponding frequency band.
The second link, labeled `Aurora link', is a duplex serial link operated at 2.5\,Gb/s implemented with the Aurora Intellectual Property (IP) provided by Xilinx \cite{Aurora}.
It is used to periodically (at $\rm 250\ MHz/2^{18} \sim 953\ Hz$) convey the 80 DDC results from each `proc' FPGA to the `split' FPGA and to bridge the IPBUS connection provided by the MTCA backplane.

An additional slow speed 12 bit DAC (AD5311), driven by the `split' FPGA, is implemented in order to generate the $\sim$500\,Hz signal that is used to modulate the frequency of the microwave synthesizer associated with the MTCA crate.
Its output is amplified and shifted to cover the -2.5\,V-+2.5\,V range.

A Phase Locked Loop (PLL) (LMK03033) uses the 10\,MHz reference clock provided via the backplane to generate the 250\,MHz system acquisition clocks (for DAC and FPGA) and the 1\,GHz clock used by the ADC.

The slow control and the data acquisition is done by a modified version of the Internet-Protocol bus (IPBUS) \cite{IPBUS}, that allows the Dynamic Host Configuration Protocol (DHCP).
The standard User Datagram Protocol (UDP) \cite{UDP}, which is a very simple Ethernet protocol, is unreliable by definition.
Consequently, the IPBUS was designed as a UDP enhancement aimed at adding a reliability mechanism.
As the IPBUS protocol is simple, it can be implemented directly in FPGA, without the need of having a Central Processing Unit.
The clocking for the control and the data readout is generated by dedicated PLL referenced by an embedded oscillator.
As required by the micro-TCA standard \cite{MTCA}, the board features a Module Management Controller (MMC).
The MMC solution used is a modified version of the one distributed by CERN \cite{MMC}.
It is composed of a micro-controller that hosts an application customized firmware.

The six FPGA used are from the same vendor: Xilinx.
The `split' FPGA  is a \mbox{XC7K70T-2FBG676C} and the `proc' are \mbox{XC7K70T-2FBG484C}.
It must be noted that each FPGA is connected to a flash memory ($\rm 16\,M \times 16\,bit$) and is configured via the Boot Parallel Interface mode (BPI).
The firmware stored in the flash memories can be updated via IPBus.

\subsubsection{Radio-frequency chain description}
The radio-frequency (RF) parts (low pass filters, mixers, modulators, ...), shown in dark orange in fig.~\ref{NIKEL_AMC_Hw}, are distributed and located close to each digital to analog converter (DAC), see fig.~\ref{PictureNikelAMC}.

\begin{figure}
\begin{center}
\includegraphics[angle=0,width=0.8\textwidth]{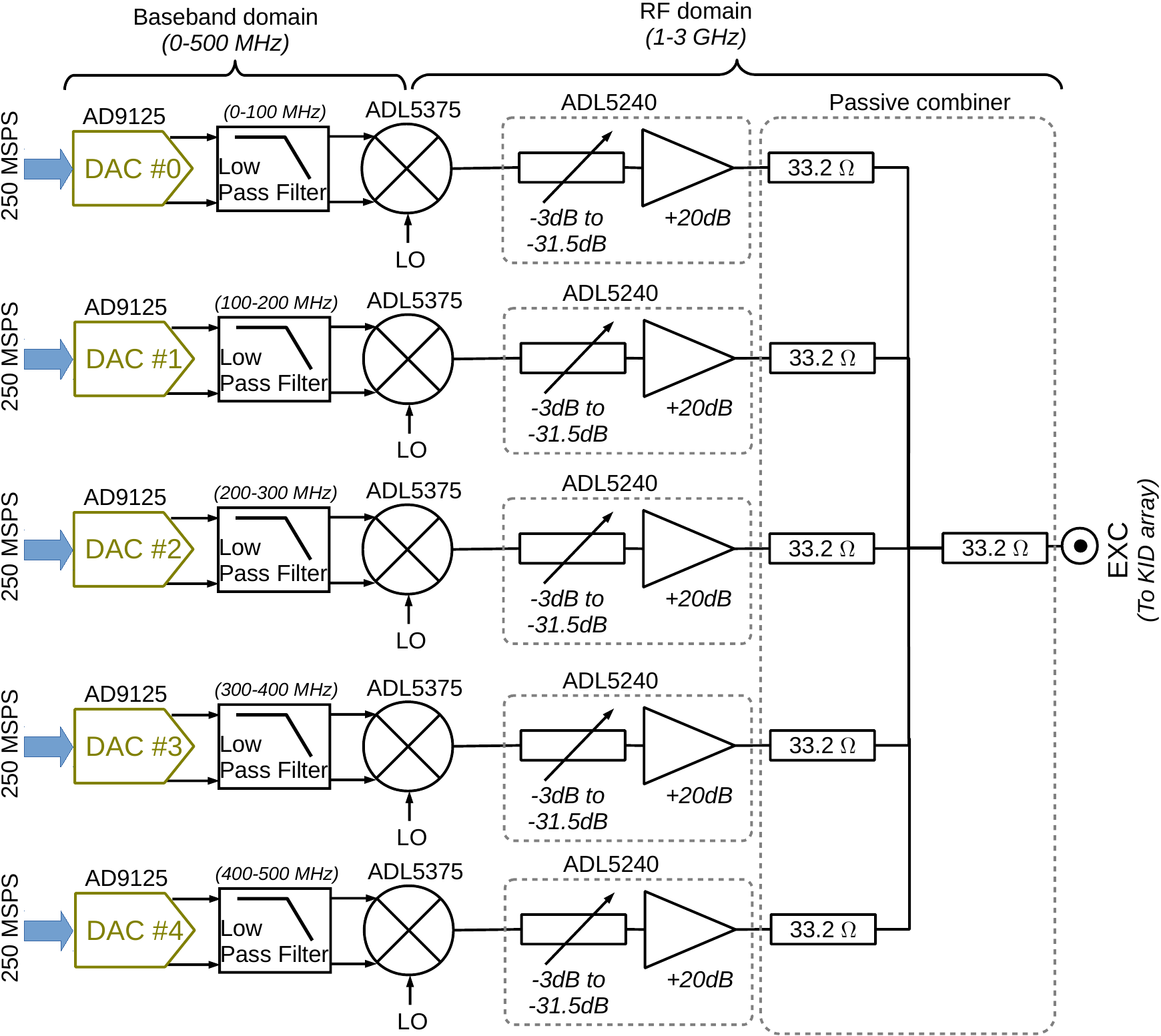}
\caption{Radio-frequency excitation front-end.
\label{DACRf}}
\end{center}
\end{figure}

At the excitation side (see fig.~\ref{DACRf}), the 1\,GHz dual output DACs are used to produce the In-phase and Quadrature signals (I/Q) for each band. 
Each of the five bands is limited to a 100\,MHz analog bandwidth.
The DAC complex modulation systems are programmed to shift the I/Q signals for band 0 to band 4 respectively in the 0-100\,MHz, 100-200\,MHz, 200-300\,MHz, 300-400\,MHz and 400-500\,MHz ranges.
The dual DAC outputs are filtered by differential 7\textsuperscript{th} order low pass filter ($\rm f_c=500\,MHz$) to remove aliasing.
In each band, the resulting signals are used to drive an IQ RF modulator (ADL5375) which performs a Single Side-Band (SSB) up-conversion.
To provide adjustment capability, the modulator output signal is then attenuated by a programmable factor (from 0\,dB to -31.5\,dB) and eventually amplified by 20\,dB.
This feature is provided by an integrated circuit ADL5240 that is controlled via IPBus through the associated `proc' FPGA.
The five adjusted excitation signals are combined through a 6-way passive combiner and sent to the KID array.
The maximum output power for a single tone generated in each band is from -10\,dBm (band 4) to -3\,dBm (band 0), for more details see section~\ref{systFreqRespSect}.

\begin{figure}
\begin{center}
\includegraphics[angle=0,width=0.7\textwidth]{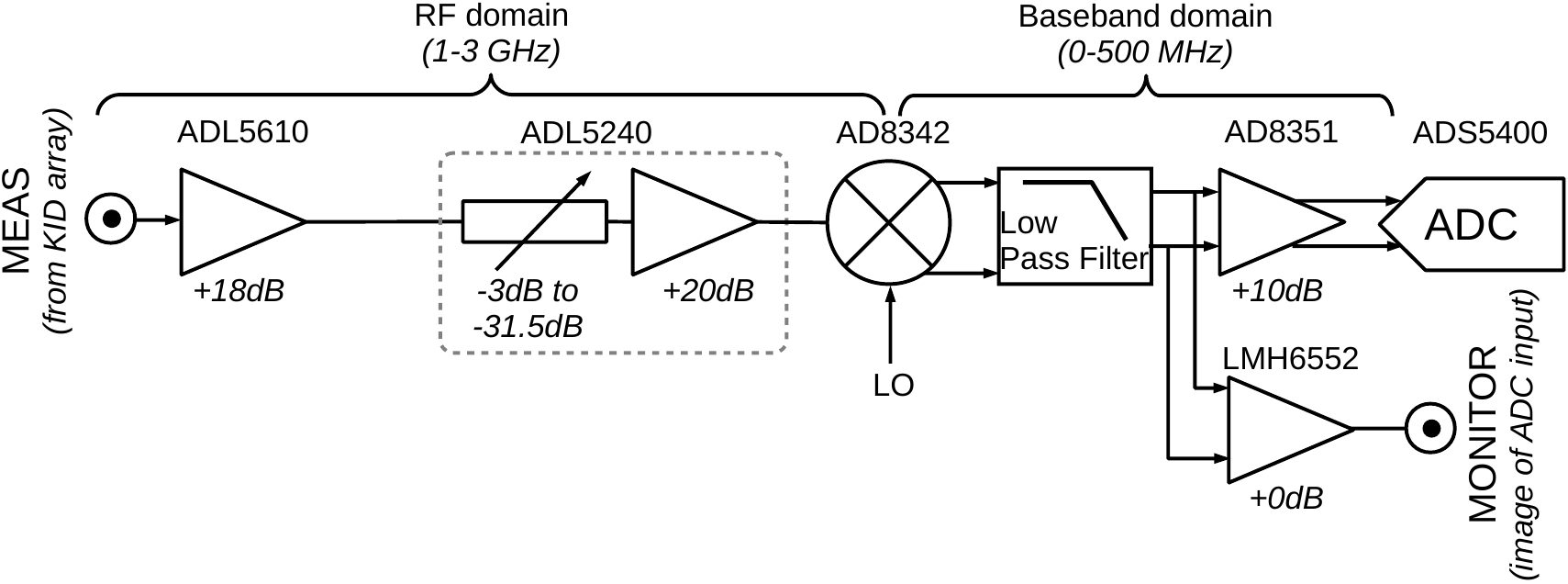}
\caption{Radio-frequency measurement front-end.
\label{ADCRf}}
\end{center}
\end{figure}

The radio-frequency measurement front-end is shown in fig.~\ref{ADCRf}.
The signal returning from the KID array is first amplified by 18\,dB by an RF amplifier (ADL5610). 
As for the excitation signal, the amplified measurement signal can be attenuated by 0\,dB to -31.5\,dB using another IPBus-controlled ADL5240 and is eventually amplified by 20\,dB.
The resulting signal is then down-converted by an AD8342 mixer (double side band)
and its differential outputs are low pass filtered by the same filter used on the excitation side to remove the unwanted frequency products.
Given the fact that the unwanted 500\,MHz sideband does not contain any carrier (see excitation scheme), only residual noise will be added to the down-converted signal band.
The low pass filter outputs are then distributed to a +10\,dB differential amplifier (AD8351) leading to the ADC and to an isolating amplifier (LMH6552) connected to a monitoring output. 

\subsection{Firmware description}

\subsubsection{FPGA `split' description}
\begin{figure}
\begin{center}
\includegraphics[angle=0,width=0.9\textwidth]{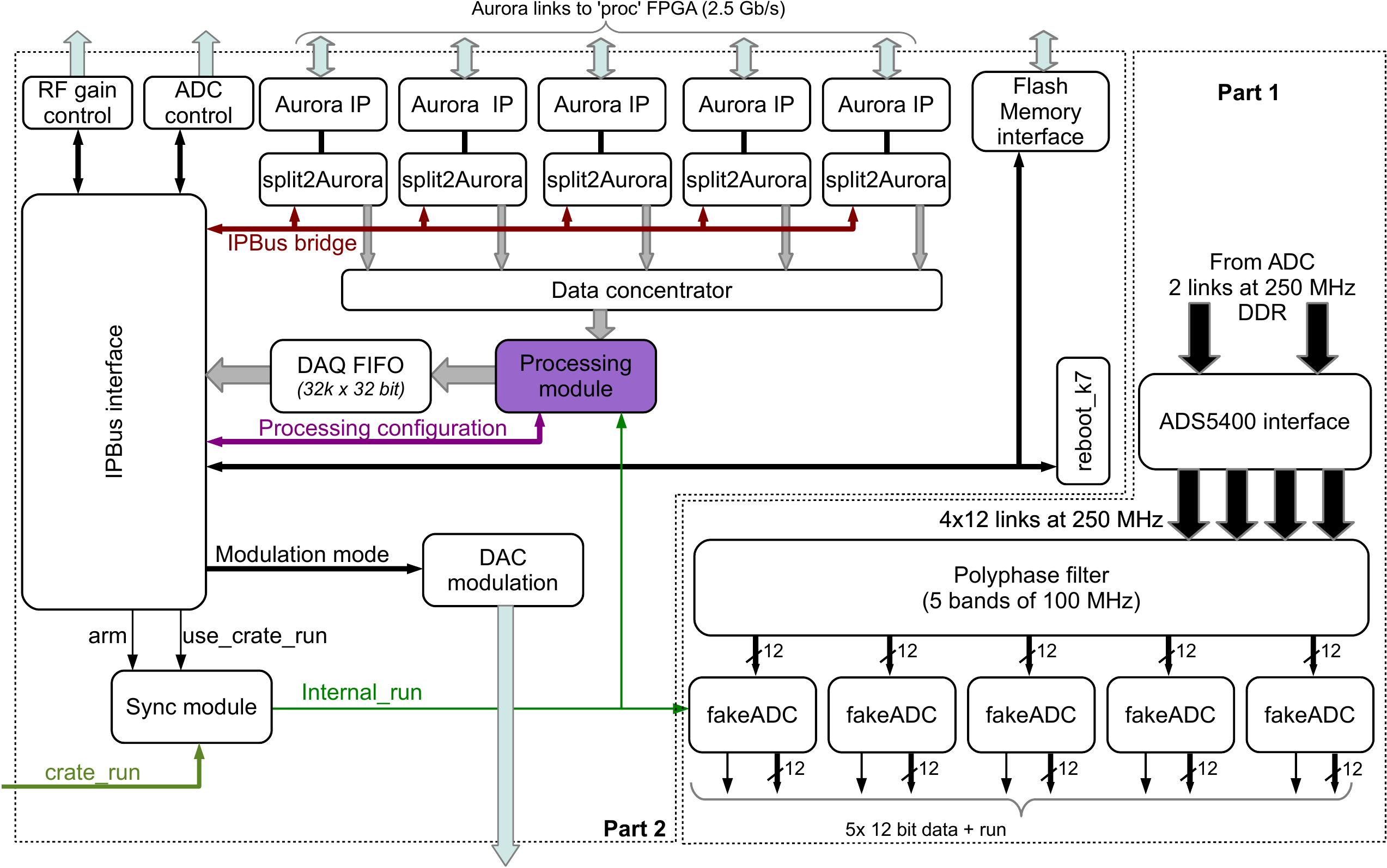}
\caption{Overview of the `split' FPGA firmware. The first part, which is the key-point of the overall design, is composed of the ADC interface, the polyphase filter bank and the five `fakeADC' outputs, each carrying its share of the bandwidth to the dedicated `proc' FPGA.
The second part consists of an IPBus interface, five Aurora transceivers, a data concentrator, a post-processing module, a large FIFO, an acquisition synchronization module, a frequency modulation controller and a flash memory interface.
\label{splitFw}}
\end{center}
\end{figure}

The `split' FPGA, shown in figure~\ref{splitFw}, contains two main parts.
The first part, which is the key-point of the overall design, is composed of the ADC interface, the polyphase filter bank and the five `fakeADC' outputs, each carrying its share of the bandwidth to the dedicated `proc' FPGA.
The implementation of this part of the design is extensively described in \cite{Bourrion2012}.
The second part consists of an IPBus interface, five Aurora transceivers, a data concentrator, a post processing module, a large FIFO, an acquisition synchronization module, a frequency modulation controller and a flash memory interface.

The IPBus interface uses a LogiCore IP to implement an Ethernet 1000BASE-X Physical Coding Sublayer/Physical Medium Attachment (PCS/PMA) with one of the Gigabit Transceiver \cite{pg047} available in the FPGA.
On the user interface side, it provides a 32 bit address and a 32 bit data bus with handshake signals.
This bus is employed to read-out the acquisition data and to configure the whole system.

The Aurora transceivers are designed to operate at a speed of 2.5\,Gb/s, which is the speed required to carry 16 bit at 125\,MHz with an 8b/10b encoding in full duplex mode.
They are composed of the `Aurora IP', a Xilinx provided IP, and the `split2Aurora'.
This latter module permits sharing the link between bridging the IPBus link to each `proc' FPGA (duplex mode) on one hand and collecting the I/Q data provided by each `proc' FPGA (simplex mode) on the other hand.
In practice, the link sharing module, allocates 8 bit for the IPBus bridging and 8 bit for readout data transport.

Every $\sim$1.05\,ms (2\textsuperscript{18} clock cycles at 250\,MHz) a 644-byte data frame is received (see section~\ref{FirmwareProcDevel}) and stored in a small reception buffer (1\,k word deep). 
Once every Aurora transceiver has received its data frame, the `data concentrator' recovers and aggregates data from each `Aurora' link and transfers them to the processing module.

The DAC modulation block is used to generate an optional modulation signal which can be a 2 or 4 values modulation signal, depending whether it is desired or not to compute the sensitivity (first derivative) and the sensitivity variation (second derivative) of the I/Q measurements.
When this block is activated, the modulation signal is modified every 2\textsuperscript{18} clock cycles, which corresponds to the `proc' FPGA integration cycle.
To ensure the modulation is synchronous with the integration performed in the DDC, the initial start of the modulation is adjustable with a resolution of 4\,ns and up to one full integration cycle. 

The `processing module' is used to compute on-line the sensitivity or the sensitivity variation of the I/Q measurements provided by the data concentrator.
It provides $k$ (1 up to 8) sets of I/Q measurements for a block of $n$ (1 up to 64) consecutive I/Q values.
The $k \times n$ set of coefficients used by the processing module is provided via the IPBUS. 
The coefficients $C$ are 4 bit signed integer values, i.e. they range from -8 to +7.
The generic equation employed to compute the resulting $I_{res}^j$ (or $Q_{res}^j$) for a given KID $j$ using $n$ samples $I^j$ (or $Q^j$) is provided in eq.~\ref{eq::1}, $shift\_val$ is adjusted to avoid computing overflow.
\begin{equation}
\begin{bmatrix}
I_{res}^j(1) & Q_{res}^j(1) \\
I_{res}^j(2) & Q_{res}^j(2) \\
... & ... \\
I_{res}^j(k) & Q_{res}^j(k) \\
\end{bmatrix}
=
\begin{bmatrix}
C_{11} & C_{12} & ... & C_{1n} \\
C_{21} & C_{22} & ... & C_{2n} \\
... & ... & ... & ... \\
C_{k1} & C_{k2} & ... & C_{kn} \\
\end{bmatrix}
\times
\begin{bmatrix}
I^j(1) & Q^j(1) \\
I^j(2) & Q^j(2) \\
I^j(3) & Q^j(3) \\
... & ... \\
I^j(n) & Q^j(n) \\
\end{bmatrix}
\times
\frac{1}{2^{shift\_val}}
\label{eq::1}
\end{equation}

\begin{figure}
\begin{center}
\includegraphics[angle=0,width=0.9\textwidth]{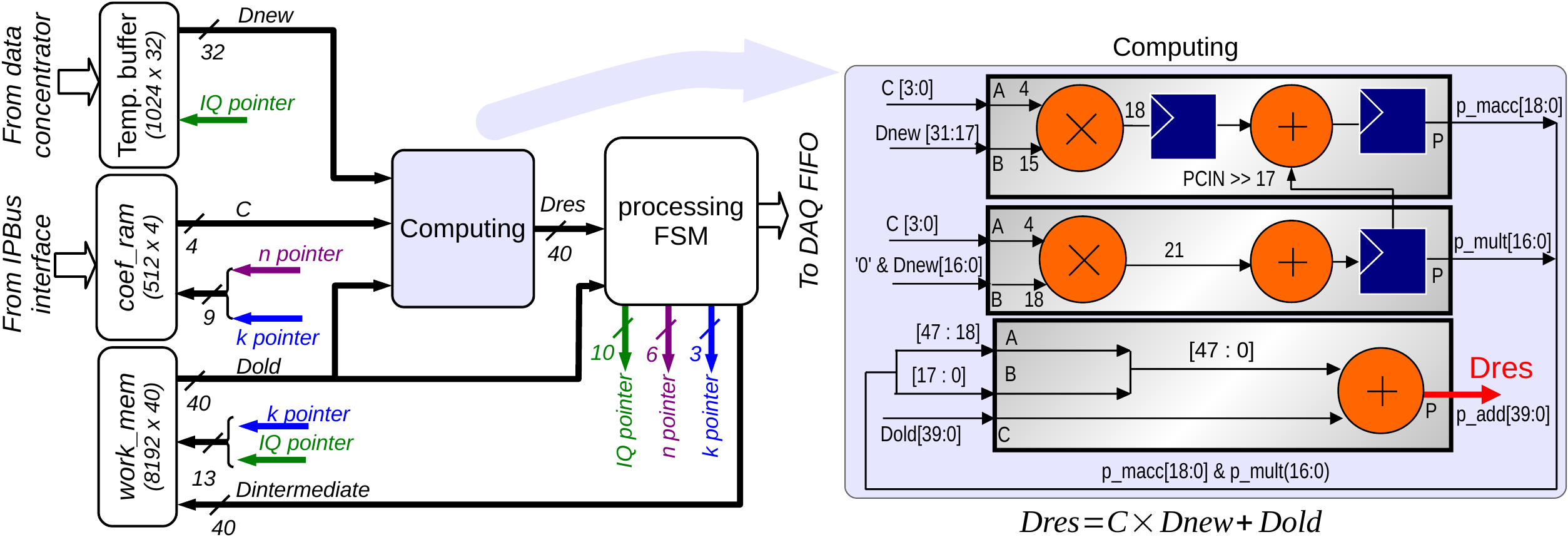}
\caption{Overview of the processing module (l.h.s) and of the computing unit (r.h.s).
\label{procModuleFw}}
\end{center}
\end{figure}

The `processing module' architecture is shown on the l.h.s of  figure~\ref{procModuleFw}. 
The data provided by the data concentrator (800 words of 32 bit) are written in a temporary buffer (1024 words of 32 bit), to cope with the computing that takes several clock cycles per operation.
The `coef\_ram' (512 words of 4 bits) stores the $n$ coefficients for all sets $k$.
The `computing block', composed of three DSP48E1 blocks, is designed to perform a multiplication between a 32 bit word (the new sample) and a 4 bit word (the coefficient) and to add a 40 bit word (the result of the previous computation).
The `work\_mem' (8192 words of 40 bits) stores the $k$ sets of intermediate $I_{res}$ and $Q_{res}$ data.
The processing finite state machine (FSM), which sequences all operations, performs two main tasks.
At first, for each I/Q data frames, it controls the different pointers, which are used to address the various memories and hence provides the coefficients, the new data and the previously computed data to the `computing block'.
Eventually, each intermediate computing result is saved in the `work\_mem'.
As a second task, when the required $n$ incoming I/Q data frames have been used for the computing, the FSM takes the data from the `work\_mem' and sends them to the `DAQ\_FIFO' (see figure~\ref{splitFw}).
This last step make the complete data frame available for data acquisition via IPBus.

As detailed above, the data rate generated by the `processing module' is a function of the number of measurements $n$ (1 up to 64) and of data sets $k$ (1 up to 8) requested.
The data rate $DR$ per readout board is expressed in eq.~\ref{eq::2}. 
\begin{equation}
DR = \frac{400 \times k \times 8}{n} \times \frac{250 \times 10^6}{2^{18}}
\label{eq::2}
\end{equation}
For NIKA2, $k$ is set at 2 (to record the dI, dQ information) and $n$ at 40 (acquisition rate of about 22\,Hz).
These settings yield a data rate of about 152\,kB/s per readout board, and consequently of about 3\,MB/s for the whole readout system.
This data rate is fully compliant with the IPBus capabilities which can reach a throughput of about  100\,MB/s in the multiple targets scenario \cite{IPBUS}.

The `split' firmware uses 219 out of 240 DSP48E1 blocks, 34502 out of 82000 slice registers, 25286 out of 41000 slice LUT and 85.5 out of 135 block RAM tiles.

\subsubsection{FPGA `proc' description}
\label{FirmwareProcDevel}
\begin{figure}[th]
\begin{center}
\includegraphics[angle=0,width=0.95\textwidth]{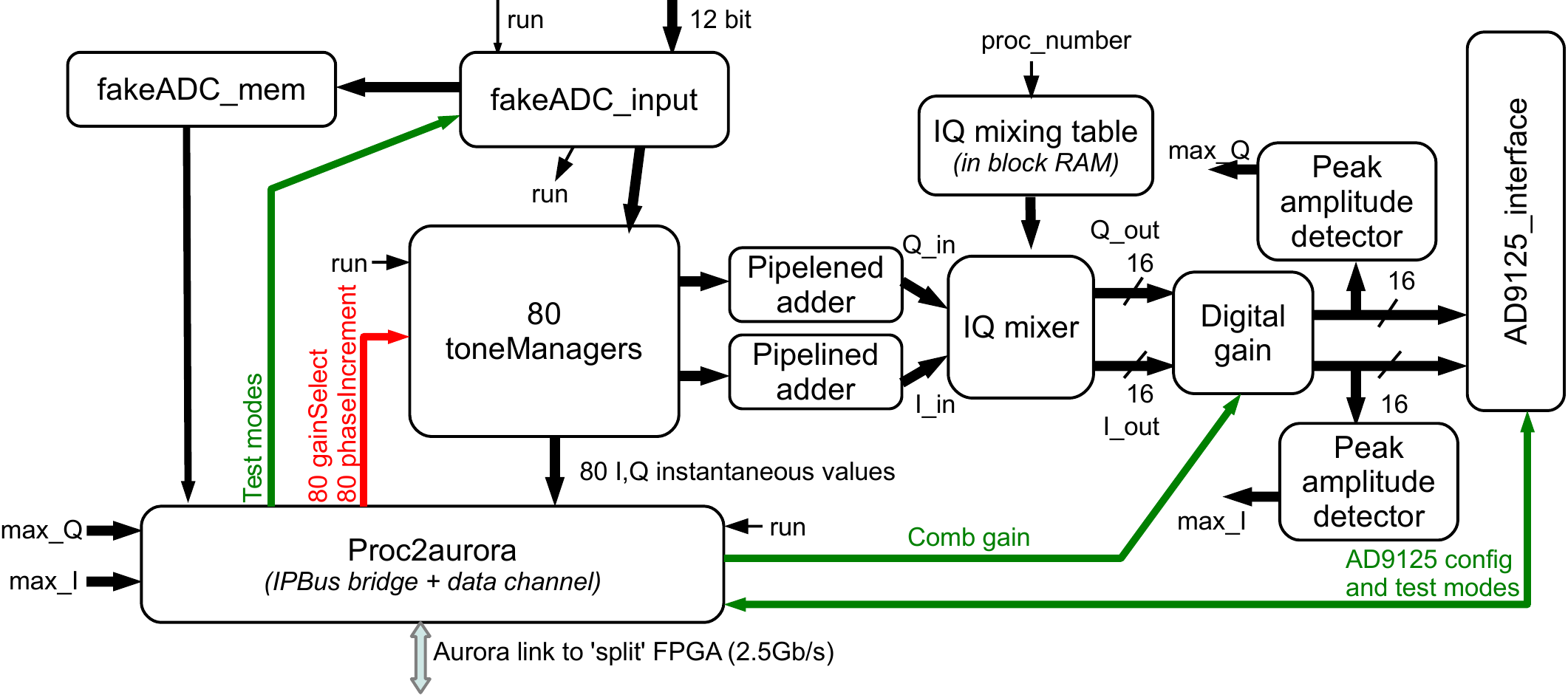}
\caption{Overview of the `proc' FPGA firmware.}
\label{FPGA_proc_nikel}
\end{center}
\end{figure}

As explained before, the processing FPGA, whose block diagram is shown in figure~\ref{FPGA_proc_nikel}, is in charge of generating the frequency comb in its bandwidth region and to perform the channelized DDC for each considered tone.

The communication between the computer and the `proc' FPGA is ensured via an IPBus connection that is bridged through the Aurora serial link.
In the `proc' FPGA, the `proc2Aurora' module directly presents the IPBus user interface composed of a 32 bit address bus and a 32 data bus with handshake signals.
Thanks to the IPBus bridging, the IPBus link can be used to set the individual phase increment values, the tone attenuations and the digital gain (`comb gain').

Configuration and test modes can also be set via this interface.
Among the provided test modes, it may be noted that it is possible to record a `fakeADC' signal snapshot of 32\,k samples in the `fakeADC\_mem'.
Moreover the DAC internal registers values can be accessed.

Given the fact that the `fakeADC' data emitted by the `split' FPGA are synchronized by the system-wide reference clock, a dedicated interface (fakeADC\_input) is used to adjust the `fakeADC' bus delay in order to compensate for the data sampling phase misalignment, and thus to guarantee stable information sampling.
The locally synchronized data are provided to the tone managers.

The 80 tone manager outputs are fed to two pipelined adders in order to construct the in-phase and quadrature versions of the frequency comb. 
Each comb version is then frequency shifted by a digital `IQ mixer' block in order to compensate the residual up-converting due to the polyphase filtering ($\rm +12.5\,MHz$) and the frequency shift due to the non optimal selection of the DAC internal modulator frequency  (see polyphase filter section in \cite{Bourrion2012} for full details).
The 16 bit digital gain is used to numerically adjust, amplify or attenuate the resulting signal before driving the DAC.
This feature is useful to adapt the signal to the ADC input range when less than 80 tones are used.
The gain $G$ applied for a given $combGain$ value is given in eq.~\ref{gainEq}.
\begin{equation}
G=\frac{combGain}{2^8}
\label{gainEq}
\end{equation}

The `proc2Aurora' block is used to transmit the DDC results through the Aurora link to the `split' FPGA for data concentration.
Along with these data, the detected peak amplitude, in absolute value, is transmitted for monitoring and to avoid DAC clipping.
Hence, the data frame is composed of $\rm 2 \times 80$ 32 bit words representing the in-phase/quadrature information.

\begin{figure}[th]
\begin{center}
\includegraphics[angle=0,width=0.75\textwidth]{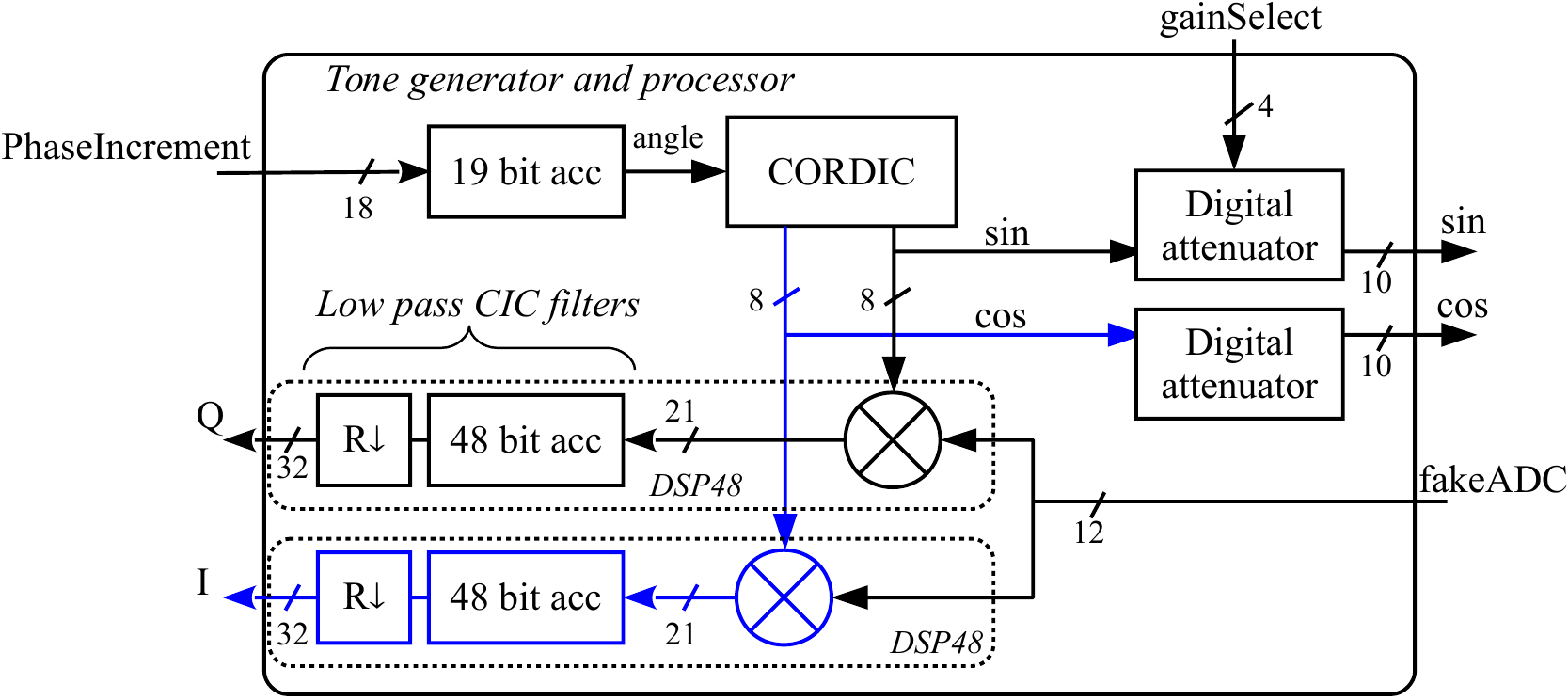}
\caption{Overview of a tone manager. The block comprises a CORDIC generator, two digital attenuators for individual tone power adjustment and a DDC implemented with DSP48 blocks.}
\label{nikel_tone_manager}
\end{center}
\end{figure}

The tone manager, which is depicted in figure~\ref{nikel_tone_manager}, features a COordinate Rotation DIgital Computer (CORDIC) \cite{Volder} block and a DDC that is composed of an I\&Q demodulator followed by a Low Pass Filter (LPF).
The LPF, which is primarily used to remove the summed frequencies component from the spectrum, also provides unwanted frequencies rejection (e.g. frequencies tuned to other pixels, white noise, \ldots).
Each CORDIC, implemented in a pipelined fashion and composed only of adders and subtracters, was designed to provide a 8 bit precision on the calculated sine and cosine values.
It uses 8 precalculated arc tangent values with 20 bit resolution.
The 19-bit phase accumulator that feeds the CORDIC is used to adjust the frequency thanks to a 18-bit Phase increment.
Hence, the frequency is tuned with a precision of $\rm 250\ MHz/2^{18} \sim 953\ Hz$.
The phase accumulators are initialized to random values at startup to avoid the coherent addition of maximum cosine or sine values of all CORDICs.

The I\&Q demodulation is performed by multiplying a copy of the ADC output by replicas of the generated sine and cosine values. For practical reasons (FPGA logic resources), the LPF is obtained by averaging $2^{18}$ data samples (bandwidth of 953\,Hz). It must be noted, that the accumulator period must be chosen as a multiple of the phase accumulator period in order to avoid beat frequency phenomena
At the end of the accumulation cycle, each tone manager transfers its I/Q data to the `proc2Aurora' interface for transmission to the `split' FPGA.

To allow individual tone power adjustment, the sine and cosine waves are passed through digital attenuators before being provided to the block output. Tones can be tuned in the range 0 to 8/8 and have a resolution of 1/8\textsuperscript{th} of the input power. 

The `proc' firmware uses 166 out of 240 DSP48E1 blocks, 47542 out of 82000 slice registers, 32677 out of 41000 slice LUT and 4 out of 135 block RAM tiles.

\section{Central clocking and synchronization board}
\begin{figure}
\begin{center}
\includegraphics[angle=0,width=0.49\textwidth]{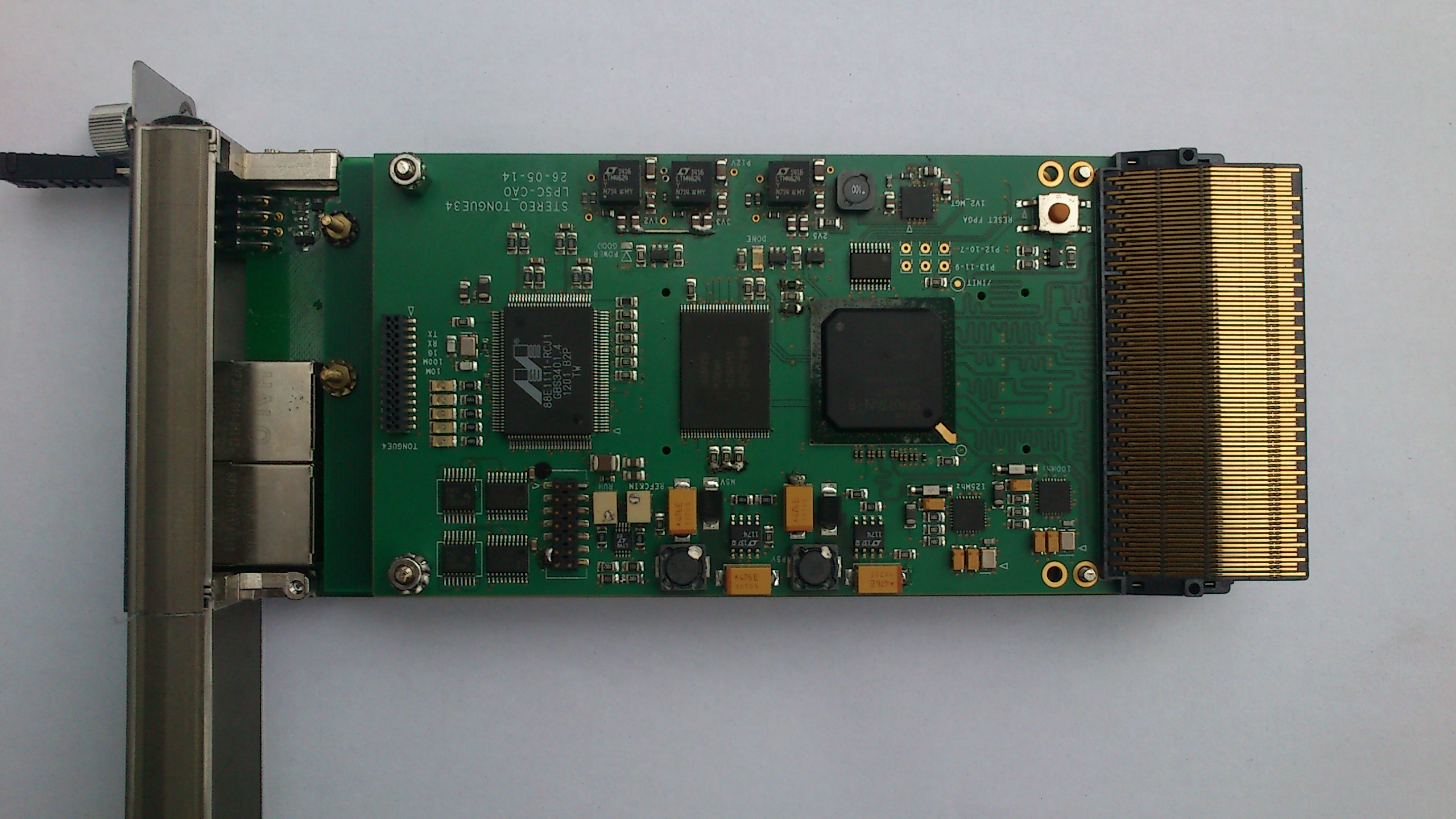}
\includegraphics[angle=0,width=0.49\textwidth]{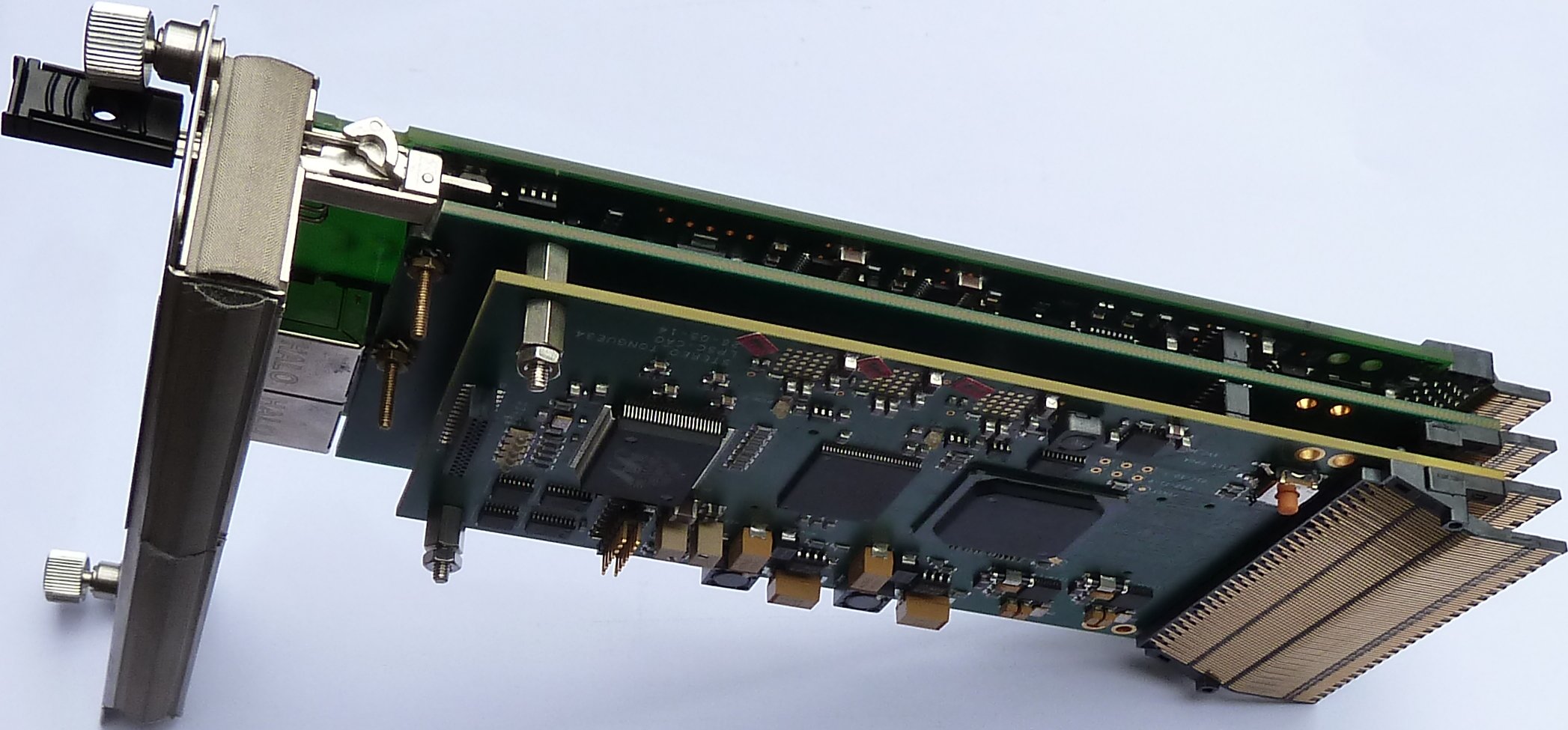}
\caption{Pictures of the two boards composing the CCSB mounted on the NAT\textregistered \  MCH. The main CCSB board is the top board in the left picture. The extension PCB providing the SMA connectivity with the front panel for the reference clock and the PPS is not mounted.\label{CCSBhw}}
\end{center}
\end{figure}

In order to benefit from the star communication architecture of the microTCA backplane, the CCSB was designed as a set of two mezzanine extension boards that are mounted on the NAT\textregistered \  MCH as shown in  fig.~\ref{CCSBhw}.
The CCSB is connected to the MCH with a connector that provides the power supply and the Ethernet connectivity.
The PCB associated with the second tongue contains only buffers for the clock tree (FCLKA) and the required connectivity for interconnecting the MCH, the tongue 2 and the main CCSB board.
The main CCSB board is equipped with an FPGA (XC6SLX45T-FGG484) which takes care of the clock selection and PPS distribution, its associated flash memory for the BPI mode, the power converters, two adjustable comparators and the connectivity with the tongue 3 and 4.
As for the NIKEL\_AMC, the flash memory on the CCSB can be updated in-situ via IPBus.

\begin{figure}
\begin{center}
\includegraphics[angle=0,width=0.8\textwidth]{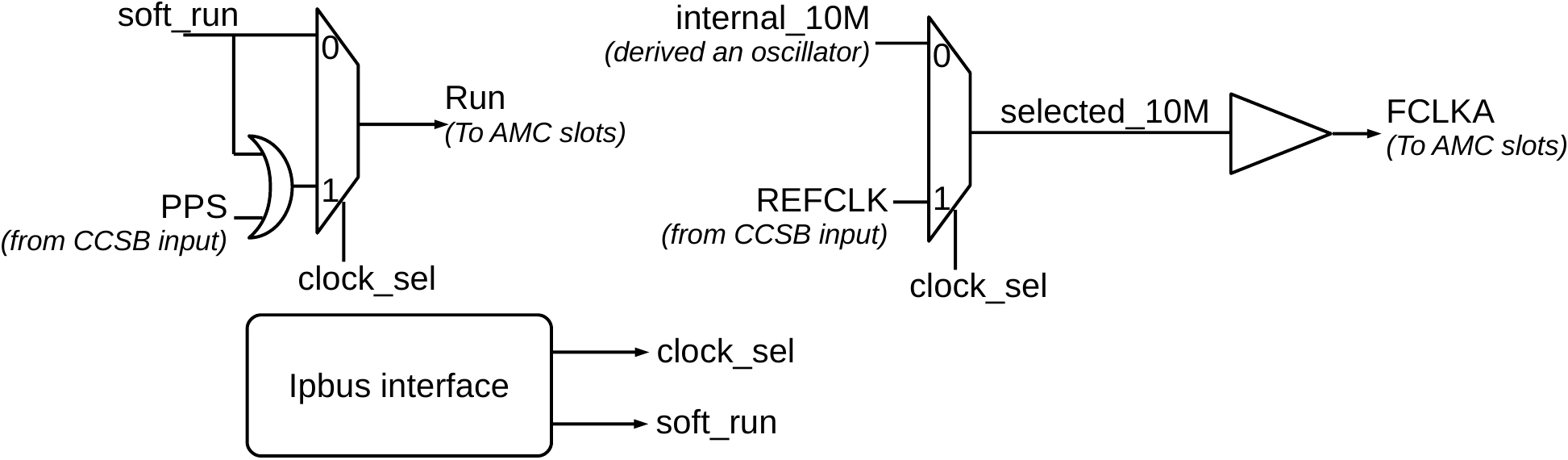}
\caption{CCSB firmware overview.
It contains the IPBus interface that provides control over the clock source (internal or external) as well as the ``run'' signal source.
\label{CCSBfw}}
\end{center}
\end{figure}

The CCSB, whose hardware was developed for an other experiment \cite{stereo}, currently holds a reduced firmware, see fig.~\ref{CCSBfw}.
The firmware contains the IPBus interface that provides control over the clock source (internal or external) as well as the ``run'' signal source.
The acquisition can be started by software command (``soft\_run'') or by the PPS.

\section{System performances}
\label{sysPerfSec}
\subsection{Power consumption}
A NIKEL\_AMC uses less than 53\,W in total power to instrument up to 400 KIDs (all tone managers activated).
The radio-frequency chain contributes for 12\,W to this power draw.

Each crate uses 120\,W for the two fan trays and 12\,W for the MCH (manufacturer specifications).
The CCSB uses about 5\,W.
Hence, a crate equipped for a 260\,GHz array requires about 550\,W while for the 150\,GHz array it requires about 330\,W.
The numbers must be adjusted to take into account the power module conversion efficiency of about 90\,\%.
The total readout electronics power for instrumenting the three NIKA2 KID arrays, for a total of 3300 detectors, is about 1590\,W.

\subsection{System transfer function}
\label{systFreqRespSect}
The NIKEL\_AMC transfer function was measured in loop-back mode, i.e. the excitation output connected to the measurement input.
This configuration is actually pretty well representative of the real working mode, where
the output signal is attenuated by roughly 30\,dB on its way to the KIDs, then boosted by 30\,dB thanks to a cryogenic Low Noise Amplifier (LNA).

\begin{figure}
\begin{center}
\includegraphics[angle=0,width=0.8\textwidth]{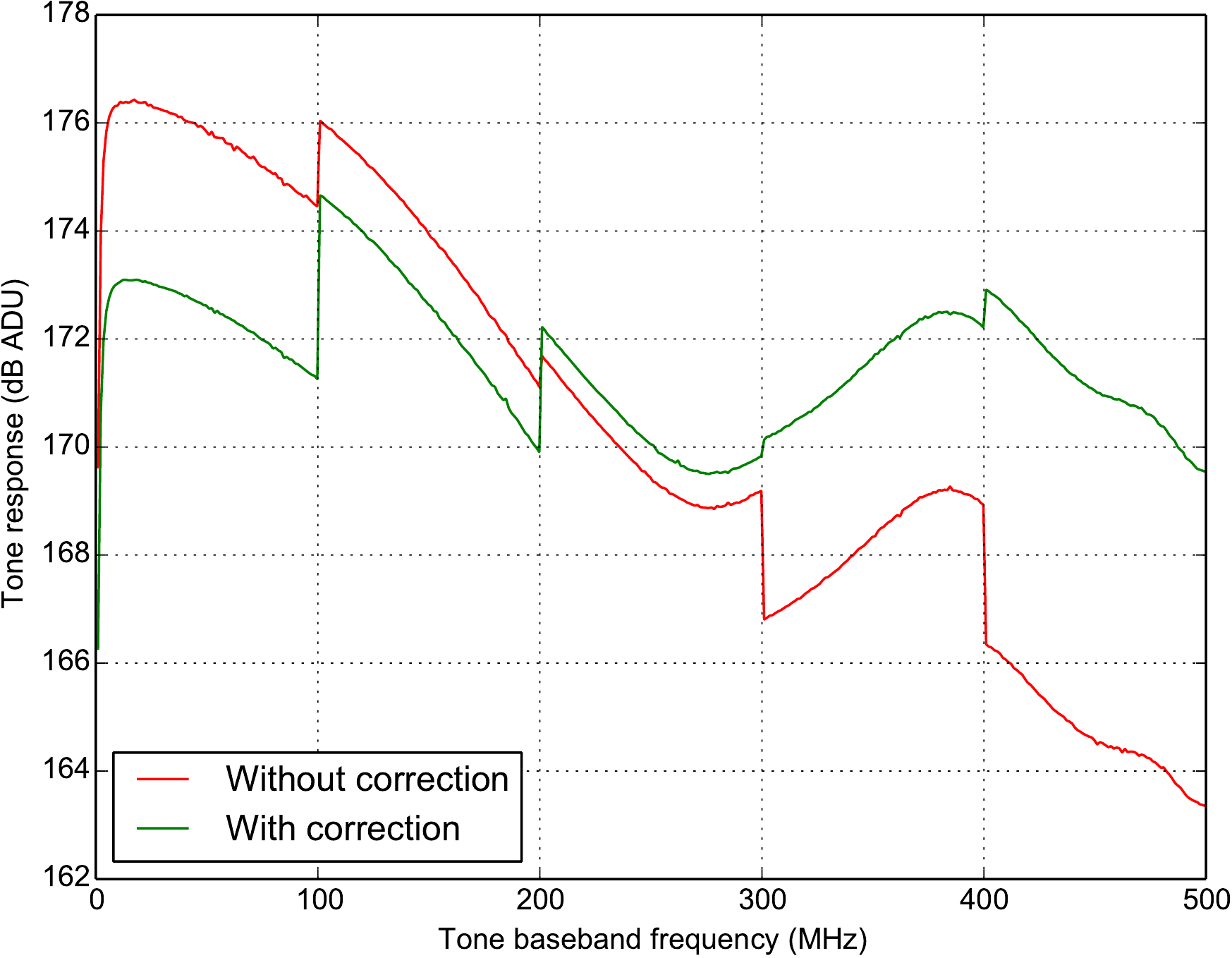}
\caption{NIKEL\_AMC transfer function with and without band amplitude compensation.
The measurements were done with a local oscillator set at 1.3\,GHz, and the electronics was tuned to generate and analyze 400 tones.
The frequencies were evenly distributed from 1\,MHz to 500\,MHz.
Each tone gain was set at 4/8\textsuperscript{th} and the digital gains ($combGain$) were set at 4096 to avoid DAC clipping.
\label{systResp}}
\end{center}
\end{figure}

Figure~\ref{systResp} shows the system transfer functions measured with a local oscillator set at 1.3\,GHz. 
NIKEL\_AMC was tuned to generate and analyze 400 tones.
The frequencies were evenly distributed from 1\,MHz to 500\,MHz.
Each tone gain was set at 4/8\textsuperscript{th} and the digital gain for each band ($combGain$) were set at 4096 to avoid DAC clipping.
In these conditions, two measurements that had operating conditions similar to those for instrumenting a 150\,GHz array, were performed.
For these, the Local Oscillator (LO) was set at 1.3\,GHz and an external 20\,dB attenuation was inserted in the loop-back path.
The ``without correction'' measurement was performed to assess the raw electronics performances, i.e. the output attenuation was tuned to 0\,dB and the measurement input attenuator was set to -12.5\,dB.
The ``with correction'' measurement, where the excitation spectrum non-uniformities were corrected, was achieved with the output attenuations set at 10\,dB, 8\,dB, 6\,dB, 3\,dB and 0\,dB, respectively from band 0 to band 4 and the measurement input attenuator was set to -16.5\,dB.
It must be noted that the measurement attenuation is set at 0\,dB when instrumenting a 260\,GHz array.

\begin{figure}
\begin{center}
\includegraphics[angle=0,width=0.48\textwidth]{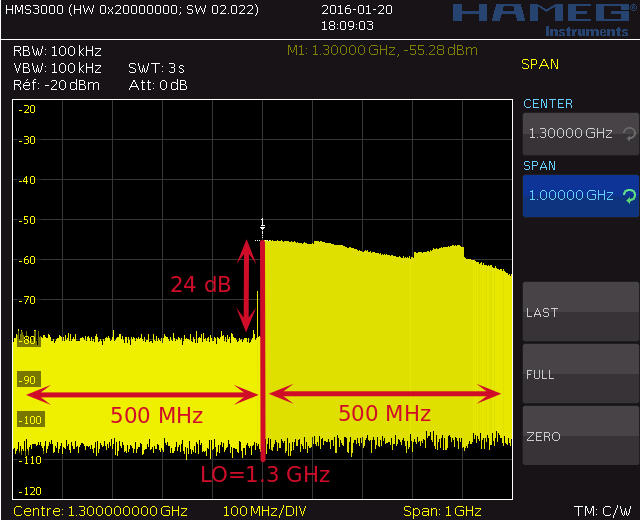}
\includegraphics[angle=0,width=0.48\textwidth]{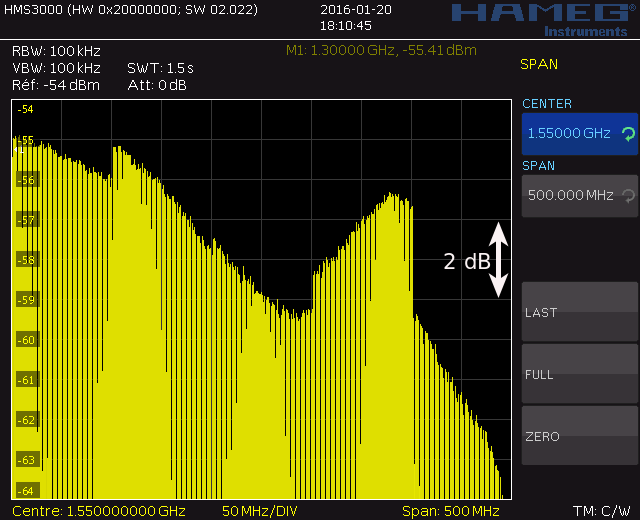}
\caption{Excitation output spectrum measurements with a LO at 1.3\,GHz. On the l.h.s capture which was taken with a frequency span of 1\,GHz, the good image frequencies rejection can be observed. On the r.h.s which was taken with a frequency span of 500\,MHz and with dilated amplitude scale, the shape of the frequency response can be seen; it is similar to the plot obtained with the loop-back measurement. 
\label{spectrumMeas}}
\end{center}
\end{figure}
As figure~\ref{systResp} shows on the uncorrected measurement, the amplitude variation is about 11\,dB over the full bandwidth.
It can be seen in figure~\ref{spectrumMeas} that theses amplitude variations are mainly due to the excitation side since the shape is very similar to the uncorrected frequency response shown in red line in figure~\ref{systResp}.
Given this fact, this non-uniformity can be easily mitigated by tuning the RF excitation attenuation and the digital comb gain individually for each of the five bands.
An example of adequate tuning using only the band gain adjustments is shown in green line in figure~\ref{systResp}.
It can be further improved by tuning the tone gain individually.
We believe these variations to be induced by two main sources detailed hereafter.

\begin{figure}
\begin{center}
\includegraphics[angle=0,width=0.8\textwidth]{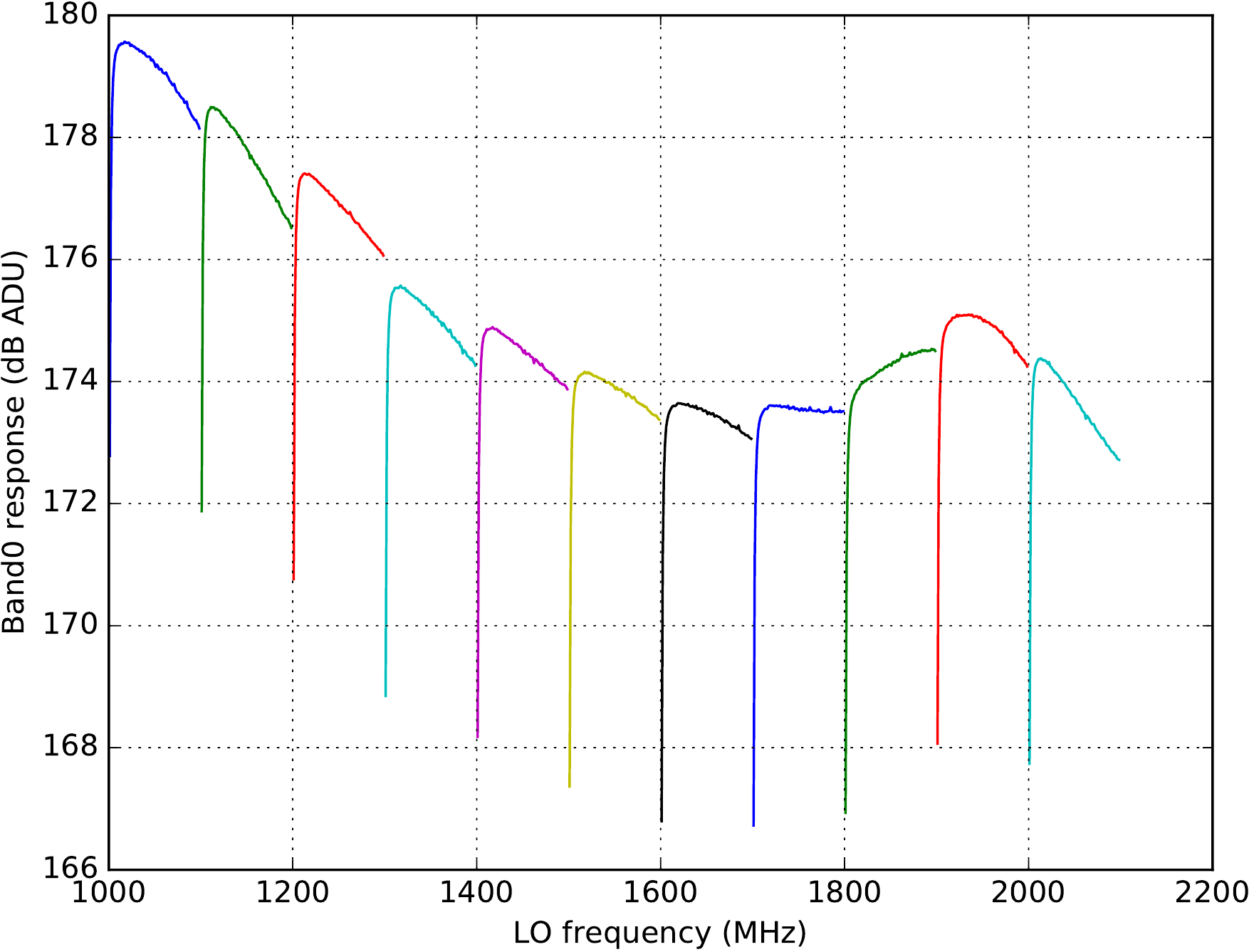}
\caption{Illustration of the PCB effects on the band 0 frequency response.
It can be seen that a variation of 500\,MHz induces an amplitude drop of 6\,dB.
The plot was obtained by varying the LO between 1\,GHz and 2\,GHz in 100\,MHz steps and recording the system response for each measurement. The 11 data sets are plotted in this figure (one color per set).
\label{band0_resp-crop}}
\end{center}
\end{figure}

The first source of amplitude discontinuity is due to the selected architecture, i.e. the five excitation bands.
Fairly long RF transmission lines (about 10\,cm) are required to connect to the passive six ways output combiner.
Even though special care has been taken on the Printed Circuit Board (PCB) layout for those micro-strip lines, the impedance is not perfectly controlled.
Indeed, PCB fabrication tolerances can induce impedance mismatch of up to 10\%, and there are parasitics and losses due to use of FR4 dielectric.
Namely, the insertion losses are estimated to be from 1\,dB to 1.5\,dB in this frequency range and for this trace length.
To illustrate these PCB effects, the frequency response of the band 0 between 1\,GHz and 2\,GHz was measured by varying the LO between 1\,GHz and 2\,GHz in 100\,MHz steps (a total of 11 measurements).
The resulting frequency response is shown in figure~\ref{band0_resp-crop}, it illustrates the frequency behavior of this line and the increasing attenuation with the increasing frequency.
It can be seen that a variation of 500\,MHz induces an amplitude drop of 6\,dB.
The  -6\,dB spike seen at the beginning of each 100\,MHz slice is due to the low frequency cut-off (AC coupling) of the ADC input which affects only band 0.

The second source of amplitude variation is due to the fact that we are dealing with broadband signals (100\,MHz at baseband and 500\,MHz in the RF domain) and for each component of the chain (mixers, amplifiers, attenuators), variations of gain, input return losses and output return losses are expected on theses bandwidths by the component manufacturers.
Perfect impedance matching between stages is therefore impossible on the whole bandwidth, and a couple of dB of gain variations is expected on the output signal.

\subsection{System noise}
To assess the system noise, we used the same set-up and the same settings as for assessing the system frequency response.
We recorded 4096 data points for the 400 KIDs in both settings, i.e. with and without correction. Adjustments to compensate the amplitude variations over the bandwidth were done in order to keep the same input power at the ADC input.
The modulation was not activated and the processing module was configured to produce averaged data for every 40 I/Q samples, i.e. output data rate about 23.84\,Hz, as in real conditions at the telescope.

\begin{figure}
\begin{center}
\includegraphics[angle=0,width=0.99\textwidth]{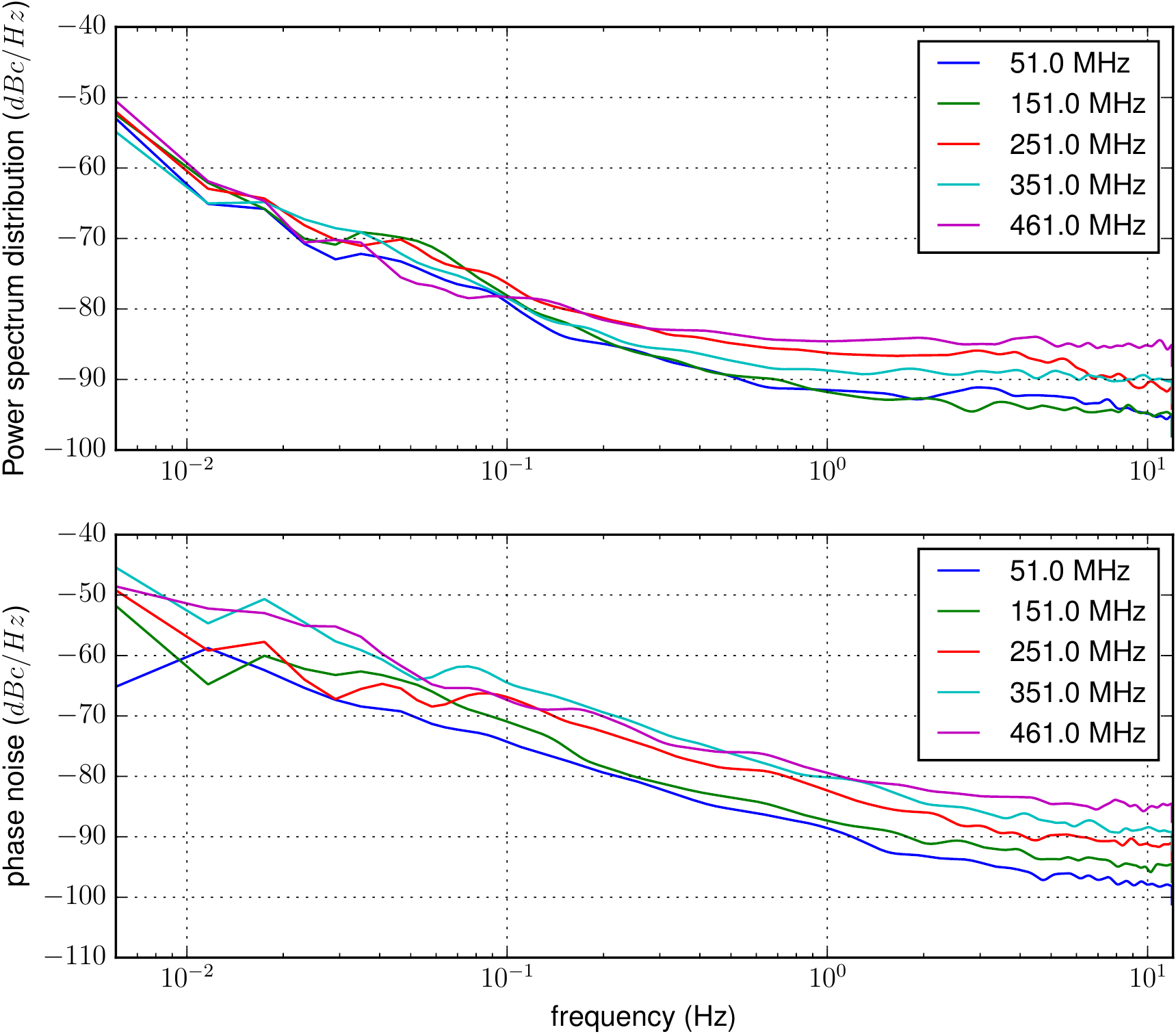}
\caption{Power spectral density plots showing amplitude (top) and phase noise (bottom) for one tone in each band. The Welch method \cite{Welch} was used to compute the noise power spectra. 
The window lengths were adapted to the frequency interval (smaller/larger window lengths at high/low frequency) to smooth the power spectra.
\label{bruit5Tones}}
\end{center}
\end{figure}
Figure~\ref{bruit5Tones} shows the power spectral densities (PSD) for the computed amplitude and phase of 5 tones; one tone per excitation band was chosen.
For each tone, the PSD is normalized to the tone amplitude.
The Welch method \cite{Welch} was used to compute the noise power spectra. 
The window lengths were adapted to the frequency interval (smaller/larger window lengths at high/low frequency) to smooth the power spectra.
These plots show that the system noise floor is reached around 1\,Hz and that it increases with the tone frequency along the 500\,MHz bandwidth of the system.
\begin{figure}
\begin{center}
\includegraphics[angle=0,width=0.99\textwidth]{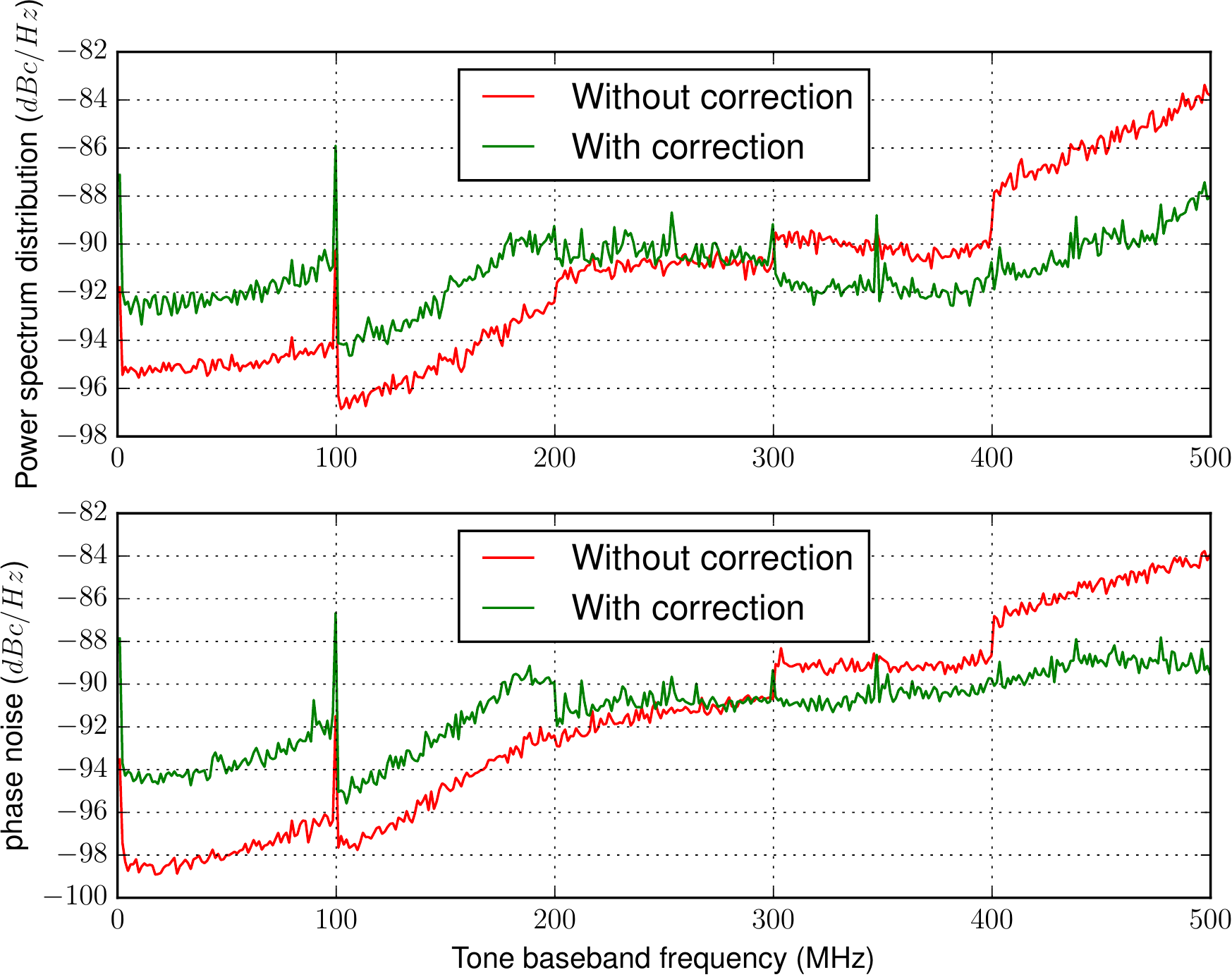}
\caption{
Average noise per tone between 10\,Hz and 11\,Hz
 without (red) and with (green) amplitude compensation.
The 400 tones are evenly distributed over the 500\,MHz bandwidth.
\label{bruit10Hz}}
\end{center}
\end{figure}

To have an overview of the phase and amplitude system noise over the full bandwidth, the average noise between 10\,Hz and 11\,Hz of each tone is plotted in figure~\ref{bruit10Hz} for both settings.
We can observe that the amplitude and phase noise have a similar shape to the system frequency response, i.e. the smaller the signal amplitude, the lower the signal to noise ratio.
Since we have maintained the same input power in the two setups, the system noise increases in the first two bands when performing amplitude compensation other the full bandwidth.
Fortunately, measurements realized with a KID array on the telescope show that with this adjustment the electronic system noise contribution remains lower than the other noise contributions (e.g. front end LNA noise) and that the whole system noise performance is improved in higher frequency bands without degrading lower bands.

We can see that for the full bandwidth, the amplitude signal to noise ratio is below -92\,dBc and the phase noise is below 30\,\textmu rad at 10\,Hz.
The spike seen at 100\,MHz is explained by digital system noise coupling.

\section{Conclusion}
The new generation of cameras for millimetre wave astronomy needs to be equipped with large detector arrays to fulfill their scientific goals. As a consequence, they 
require new electronic readout systems enabling large multiplexing factors with reduced power consumption and minimal physical space.
This is the case for the new generation NIKA2 camera that is equipped with  3 arrays of KIDs for a total 3300 pixels and has been recently installed at the IRAM 30 m telescope. 
We developed a dedicated electronic system for the NIKA2 camera that fully achieved the required performances as measured in laboratory testing.
The NIKA2 readout system is composed of 20 readout boards, {\it NIKEL\_AMC}. Each readout electronic board is able to instrument a 400 KIDs resonator feed-line 
in a bandwidth of 500\,MHz and in the frequency range from 1.0\,GHz to 2.4\,GHz with a phase noise lower than 35\ \textmu rad at 10\,Hz.
A more detailed summary of the readout electronics characteristics for the NIKA2 instrument is given in table~\ref{specTable}.

The full readout system also fulfills the system requirements for installation at the IRAM 30\,m telescope.
In particular, it uses less than 1600\,W to instrument the 3300 KIDs used by the NIKA2 camera and fits in three $\rm 19`` \times 7\,U$ crates, 
hence occupying a minimal space in the telescope cabin.

These electronics have been installed at the telescope and first measurements on the sky confirm the performances observed in the laboratory 
as shown in a companion paper describing the NIKA2 instrument \cite{catalanonika2}. The overall performances of the NIKA2 instrument are in line with expectations
making it a state-of-the-art instrument for millimetre wave astronomy. By meeting the requirements we have ensured that the readout electronics are not limiting the sensitivity
of the instrument. This is important as KIDs are expected to be photon noise limited at these frequencies \cite{Monfardini2011}.

The developments presented in this paper open the way for future KID-based cameras, which will impose even more stringent constraints and challenges in the readout electronics.
For example, these are higher multiplexing factors (higher analog bandwidth and higher DDC count) as well as lower power and radiation tolerance for space missions.

\begin{table}
\begin{center}
\begin{tabular}{r|cc}
  & 150\,GHz array & 260\,GHz arrays \\
\hline
\hline
Number of arrays & 1 & 2\\
average KIDs per feed-line & 255 & 142.5\\
Board count & 4 & 16\\
Power consumption & 370\,W & 1220\,W\\
Tone tuning resolution & 953\,Hz & 953\,Hz\\
Frequency range & 1.3-1.8\,GHz & 1.9-2.4\,GHz\\
Phase noise (at 10\,Hz) & \multicolumn{2}{c}{35\,\textmu rad}\\
Noise floor (at 10\,Hz) & \multicolumn{2}{c}{-90\,dBc} \\
\end{tabular}
\caption{Summary of the readout electronics characteristics for the NIKA2 instrument.\label{specTable}}
\end{center}
\end{table}
\section*{Acknowledgements}
This NIKA collaboration has been funded by the ANR-12-BS05-0007.
This work has been partially funded by the Foundation Nanoscience Grenoble, the ANR under the contracts ``MKIDS'' and ``NIKA''. 
This work has been partially supported by the LabEx FOCUS ANR-11-LABX-0013.
This work has benefited from the support of the European Research Council Advanced Grant ORISTARS under the European Union's Seventh Framework Programme (Grant Agreement no. 291294).
Thanks to Charles Romero and Samuel Leclercq for proof-reading the manuscript.


\end{document}